\documentclass[12pt]{article}
\usepackage{latexsym,epsf}
\newcommand{\be}{\begin{equation}}
\newcommand{\ee}{\end{equation}}
\newcommand{\bea}{\begin{eqnarray}}
\newcommand{\eea}{\end{eqnarray}}
\newcommand{\pkt}{\; .}
\newcommand{\kma}{\; ,}

\newcommand{\re}{{\rm Re}}

\newcommand{\sub}[1]{_{\rm #1}}
\newcommand{\abo}{\partial^\mu}
\newcommand{\abu}{\partial_\mu}
\newcommand{\pfi}{\varphi^a}
\newcommand{\eio}{a^{\mu a}}
\newcommand{\eiu}{a^a_\mu}
\newcommand{\pot}[1]{\frac{\lambda}{#1}}
\newcommand{\eten}{\epsilon^{abc}}

\newcommand{\intk}{\int\!\frac{{\rm d^3}k }{(2\pi)^3}\,}

\newcommand{\cab}{c_{\alpha\beta}}

\newcommand{\dt}{\frac{\rm d}{{\rm d}t}}
\newcommand{\dtt}{\frac{\rm d^2}{{\rm d}t^2}}

\newcommand{\oma}{\omega_{a0}}
\newcommand{\omp}{\omega_{\varphi0}}

\begin{document}
\begin{titlepage}
\begin{flushright}

DO-TH-96/27\\
December 1996
\end{flushright}

\vspace{20mm}
\begin{center}
{\Large \bf
Nonequilibrium dynamics:\\ preheating in the SU(2) Higgs model  }

\vspace{10mm}

{\large  J\"urgen Baacke\footnote{
e-mail:~baacke@physik.uni-dortmund.de}, Katrin Heitmann
\footnote{e-mail:~heitmann@hal1.physik.uni-dortmund.de} and
Carsten P\"atzold
\footnote{e-mail:~paetzold@hal1.physik.uni-dortmund.de}} \\
{\large Institut f\"ur Physik, Universit\"at Dortmund} \\
{\large D - 44221 Dortmund , Germany}
\\
\vspace{8mm}
\bf{Abstract}
\end{center}
The term `preheating' has been introduced recently to denote
the process in which energy is transferred from a classical
 inflaton field into fluctuating field (particle)
 degrees of freedom  without generating yet a real thermal ensemble.
The models considered up to now include, besides the inflaton field,
scalar or fermionic fluctuations.
On the other hand the typical ingredient of an inflationary scenario
is a nonabelian spontaneously
broken gauge theory. So the formalism should also be developed
to include gauge field fluctuations excited by the inflaton or
Higgs field.
We have chosen here, as the simplest nonabelian example,
the SU(2) Higgs model. We consider the model at temperature zero.
From the technical point of view we generalize an analytical and
numerical renormalized formalism 
developed by us recently to  coupled channnel
systems. We use the 't Hooft-Feynman gauge and 
dimensional regularization.
We present some numerical results but reserve a more
exhaustive discussion of solutions within the
paramter space of two couplings and the initial
value of the Higgs field to a future publication. 

\end{titlepage}
\section{Introduction} \label{intro}
Nonequilibrium dynamics in quantum field theory within 
the closed time path (CTP) formalism has become
recently a fastly developing area of research. Pioneering work
by  \cite{CalHu}  has been followed by applications to inflationary
cosmology  \cite{BVHLS,BAVHL,Son} 
and to the hadronic phase transition, especially
the possibility of formation of chiral condensates \cite{BoVeHo,CKMP}. 
With the increase of the experimental lower bound 
of the Higgs mass the electroweak phase transition 
may be second order and could then become a realistic
field of application as well. 

The typical numerical computations - or `experiments' -
in this new field have included up to now, in addition to the 
`Higgs' of `inflaton' field, scalar and fermionic fluctuations.
The cosmological application has been prepared \cite{Ve,BVH,BCVHSS} by
considering the nonequilibrium time development in a constantly
curved space. Besides Friedmann - Robertson - Walker cosmology 
another typical ingredient of inflationary scenarios 
or of  the cosmological electroweak phase transition
are nonabelian spontaneously
broken gauge theories. So the formalism should also be developed
to include gauge field fluctuations.
It is the aim of our present work to describe the analytical
and computational tools for such applications.

The analytical part includes the formulation of the 
theory and renormalization. As a convenient gauge, 
used extensively in perturbative and nonperturbative
calculations in the electroweak theory, we have chosen
the `t Hooft-Feynman background gauge. 
We will not be able
to discuss gauge invariance, especially since
already the Ansatz for the classical Higgs field implies
a choice of gauge. In the - closely related -
formulation of the effective potential it has been proposed
recently \cite{BFH} to use the absolute value squared of the
Higgs field as a gauge invariant order parameter, a choice that
merits consideration also in the present context.

The renormalization conditions were chosen such the tree
effective potential remain unchanged around the minimum
corresponding to the broken Higgs phase.
The renormalization has been based on dimensional regularization.

The numerical computation and the analytical one are both
contained in a common scheme that we have proposed 
recently \cite{BHP} for  such nonequilibrium processes. 
The main characteristics of the method are:
(i) a clean separation of the divergent and finite  parts 
of the fluctuation integrals in close relation to
CTP perturbation theory;
(ii) analytic computation of the leading order contributions
using standard covariant regularization schemes;
(iii) numerical computation of the finite parts avoiding
small differences of large numbers - the leading orders are
not subtracted from the integrand but omitted from the outset.
A fourth property has been mentioned in \cite{BHP} but not yet used:
the fact that the method can be extended easily to coupled channel
systems. This application of the method will be demonstrated in this
paper within the context of the SU(2) Higgs model.
 
The plan of this work is as follows:
in section \ref{basics} we recall the basic definitions and relations;
in section \ref{equ_mot}
 we present the one-loop nonlinear relaxation equations;
we prepare the regularization in section 
\ref{pertex} by expanding the
fluctuation modes in orders of the vertex function governed by the
classical field and by deriving the large momentum behaviour
of the first terms; regularization is then straightforward, 
the renormalization requires some algebra, both are presented in
section \ref{renorm}; the numerical computation is discussed 
in section;
we conclude in section \ref{conclus}
with a discussion of the numerical results
and an outlook to more realistic and more general applications of
the method.  

\section{Basic relations} \label{basics}
The SU(2) Higgs model is defined by the Lagrangean density
\be
{\cal L}=-\frac{1}{4}F_{\mu\nu}^aF^{a\mu\nu}
+\frac{1}{2}(D_\mu\Phi)^\dagger
(D^\mu\Phi)-V(\Phi^\dagger\Phi) \pkt
\ee
Here $\Phi$ denotes a complex Higgs doublet.
The covariant derivative is defined as
\be
D_\mu = \partial_\mu-i\frac{g}{2}A_\mu^a\tau^a
\ee
and the corresponding field strength tensor is given by
\be
F_{\mu\nu}^a
=\partial_\mu{A_\nu^a}-\partial_\nu{A_\mu^a}+g\epsilon^{abc}
{A_\mu^b}{A_\nu^c} \pkt
\ee
We write the  Higgs potential in the form
\be \label{tree}
V(\Phi^\dagger\Phi)= \frac{1}{2}\mu^2 \Phi^\dagger\Phi
+\frac{\lambda}{4} (\Phi^\dagger\Phi)^2\pkt
\ee
In the SU(2) Higgs model
$\mu^2=-2m_h^2$ where $m_h$ is the mass of the Higgs field.
In the case of unbroken symmetry - not considered here -
the mass term would be defined by $\mu^2 = m_s^2$ , where
$m_s$ is the mass of the complex scalar doublet.
We denote the "classical" or "inflaton" field as $H_0$; it
is supposed to be constant in space and depends only on the time
$t$. We will treat here the fluctuations in one-loop
order, generalizations to the $1/N_c$ expansion and
to the Hartree approximation are straightforward;
they are discussed e.g. in \cite{CoHaKlMo,BVHS}. We therefore
decompose the Higgs field into the inflaton and the
fluctuation parts as
\be
\Phi(x) = (H_0(t)+h(x)+i\tau^a\varphi^a(x))\left(
\begin{array}{c}0\\1\end{array}\right)  \pkt
\ee
Here $h(x)$ is the isoscalar Higgs field fluctuation, the
isovector fluctuations $\varphi^a(x)$, the ``would-be
Goldstone fields'' will couple to the gauge fields.
Since there is no classical gauge field, the gauge field
reduces to
\be
F_{\mu\nu}^a
=\partial_\mu{a_\nu^a}-\partial_\nu{a_\mu^a}
+g\epsilon^{abc}{a_\mu^b}{a_\nu^c} 
\ee
in terms of the fluctuations $a_\mu^a(x)$.
Furthermore we have to introduce
a gauge fixing and Faddeev-Popov Lagrangean. It is convenient
to use the 't Hooft- Feynman gauge with the gauge condition
\be
F^a(a_\mu,\varphi)=\abu\eio+\frac{1}{2}g(H_0+h)\pfi \kma
\ee
and the gauge fixing term
\be
{\cal L}\sub{gf}=-\frac{1}{2}F^a F^a \pkt
\ee
The corresponding Faddeev-Popov Lagrangean reads
\bea
{\cal L}\sub{FP}&=& \eta^{\dagger a}
\left(-\Box-\frac{g^2}{4} (H_0+h)^2\right)
\eta^a\nonumber\\
&&+\eta^{\dagger a}\left(g\eten a^{\mu c}
\abu+g\eten \abu a^{\mu c}
+\frac{g^2}{4}\varphi^b\pfi+\frac{g^2}{4}
\eten(H_0+h)\varphi^c\right)\eta^b.
\eea
The complete Lagrangean  is then given by
\be
{\cal L}\sub{tot}={\cal L} + {\cal L}\sub{gf} + {\cal L}\sub{FP}
= {\cal L}\sub{0} +{\cal L}\sub{I} \pkt
\ee
The propagators and vertices can be read off from
the free Lagrangean
\bea
\label{freel}
{\cal L}\sub{0}&=&-\frac{1}{2}\abu a_\nu^a\abo a^{\nu a}+
\frac{g^2}{8}v^2\eiu\eio\nonumber\\
&&+\frac{1}{2}\partial_0
H_0\partial^0 H_0-\pot{4}(H_0^2-v^2)^2\nonumber\\
&&+\frac{1}{2}\abu h\abo h-\lambda  v^2 h^2\nonumber\\
&&+\frac{1}{2}\abu\pfi\abo\pfi-\frac{g^2}{8}v^2\pfi\pfi\nonumber\\
&&+\abu\eta^{\dagger a}\abo\eta^a-\frac{g^2}{4}v^2
\eta^{\dagger a}\eta^a\\[4mm]
\eea
and the interaction Lagrangean
\bea
{\cal L}\sub{I}&=& -g \epsilon^{abc}a^b_\mu
a^c_\nu\partial^\mu a^{\nu a}-
\frac{g^2}{4}\epsilon^{abc}a^b_\mu a^c_\nu \epsilon^{ade}a^{\mu d} 
a^{\nu e}\nonumber\\
&&+\partial_0 H_0\partial^0 h 
+g(\partial_0 H_0) a^{0 a} \pfi+g(\abu h)\eio\pfi
+\frac{g}{2}\epsilon^{abc}(\abu\pfi) a^{\mu b}\varphi^c \nonumber\\
&&+\frac{g^2}{8} (H_0^2-v^2)\eio\eiu
+\frac{g^2}{4}H_0 h \eio\eiu+\frac{g^2}{8}
h^2\eio\eiu+\frac{g^2}{8}\eio\eiu\varphi^b\varphi^b \nonumber\\
&&-\lambda h H_0^3-\frac{3}{2}\lambda h^2(H_0^2-v^2)-
\pot{2}(H_0^2-v^2)\pfi\pfi-\lambda h^3 H_0\nonumber\\
&&-\lambda hH_0\pfi\pfi
-\pot{4} h^4-\pot{2} h^2\pfi\pfi-\pot{4}\pfi\pfi\varphi^b\varphi^b
+\lambda v^2 h H_0\nonumber\\
&&-\frac{g^2}{8}(H_0^2-v^2)\pfi\pfi
-\frac{g^2}{8}h^2\pfi\pfi-\frac{g^2}{4}H_0h\pfi\pfi\nonumber\\
&&-\frac{g^2}{2}H_0 h
\eta^{\dagger a}\eta^a-\frac{g^2}{4}h^2\eta^{\dagger a}
\eta^a-\frac{g^2}{4}\eta^{\dagger a}(H_0^2-v^2)\eta^a\nonumber\\
&&+\eten\left(g\eta^{\dagger a}\eta^b\abu a^{\mu c}+g
\eta^{\dagger a}(\abu\eta^b) a^{\mu c}+
\frac{g^2}{4}(H_0+h)\eta^{\dagger a}\eta^b
\varphi^c\right)\nonumber\\
&&+\frac{g^2}{4}\pfi\varphi^b\eta^{\dagger a}\eta^b \pkt
\eea

\section{Equations of motion}
\label{equ_mot}
The formalism of nonequilibrium dynamics in quantum field theory
and the use of the tadpole method \cite{We}
have been presented or reviewed recently by various authors
\cite{BdeVH}. We give here just the one-loop equations of motion
which are obtained from this formalism.

The basic graph from which the equation of motion of the
inflaton field is derived is depicted in Fig. 1.
The propagators are here the exact propagators in the
inflaton background field. In contrast to the free
propagators some of them involve coupled channels. We
adapt the notation to this general case by using for the
Green functions the notation $G^{++}_{jl}$.
They will be defined below. The lower latin subscripts
correspond to the different fluctuating fields
$h,\varphi,a^0,a^i$ and $\eta$ introduced in the
previous section. For all one loop integrals the
contributions of the space components of the gauge fields
and the Faddeev-Popov fields will combine since they involve
the same propagators (Green functions). We will therefore
introduce the set of subscripts $h,\bot,a$ and $\varphi$ for
the isoscalar component of the Higgs field, the `transverse'
components of the gauge fields (i.e.the combination of their
space components and the Faddev-Popov fields), the
time components of the gauge fields and the isoscalar
components of the Higgs field, respectively.
 
The vertices are realized by a matrix of
 vertex operators $Q_{jl}(t)$ 
which has the following nonvanishing components
\bea
Q_{hh}(t)&=&3 \lambda H_0(t) \nonumber \\
Q_{\bot\bot}(t)&=& \frac{3}{4}g^2 H_0(t) \nonumber \\
Q_{aa}(t)&=&- \frac{3}{4}g^2 H_0(t) \\
Q_{\varphi\varphi}(t)&=& 3(\lambda+\frac{g^2}{4})H_0(t)\nonumber  \\
Q_{\varphi a}(t)&=&Q_{a\varphi}(t)= -\frac{3}{2}g\dt \nonumber \pkt
\eea
With these notations  
the differential equation for the inflaton field
reads
\be
\ddot{H_0}(t)+\lambda(H_0^2(t)-v^2)H_0(t)
-i\sum_{jl}Q_{jl}(t)G^{++}_{lj}(t,t)=0\pkt
\ee
The propagators $G^{++}_{jl}$ are the usual
time ordered Green functions. 
We have omitted the spatial variables $\vec x$ and $\vec x'$ since
the Green functions are taken at $\vec x = \vec x'$ and, due to
translation invariance, are then independent of $\vec x$.
These Green functions at $\vec x =\vec x'$ can 
be written in terms of Fourier components as
\be
G_{jl}^{++}(t,t')=\intk G_{jl}^{++}(\vec k,t,t') \pkt
\ee
The Green functions $G_{jl}^{++}(\vec k,t,t')$ for momentum $\vec k$ 
are obtained in the usual way from the mode
functions for the various fluctuations in the time dependent
background field $H_0(t)$. We discuss briefly the different
channels.  

For the isoscalar part of the Higgs field
$G_{hh}^{++}(\vec k,t,t')$ is expressed as
\bea
G_{hh}^{++}(\vec k,t,t')&=
&\frac{i}{2\omega_{h0}(\vec k)}U_{h}(\vec k,t)U^*_{h}(\vec k,t')
\theta(t-t')\nonumber \\
&&+
\frac{i}{2\omega_{h0}(\vec k)}U_{h}(\vec k,t')
U^*_{h}(\vec k,t)\theta(t'-t),
\eea
where the mode functions $U_{h}(\vec k,t)$ satisfy the
differential equation
\be
\left[\frac{\rm d^2}{\rm dt^2}+
\omega^2_h(\vec k,t)\right]U_h(\vec k,t)=0\pkt
\ee
The mode frequency $ \omega_{h}(\vec k,t)$
is defined by
\be
\omega^2_h(\vec k,t)=\vec k^2+m_h^2+3\lambda
(H_0(t)^2-v^2)
\ee
and the `frequency at t=0' is given by
\be
\omega_{h0}(\vec k)=\omega_h(\vec k,0)\pkt
\ee
The initial conditions are
\begin{eqnarray} \label{incon1}
U_h(\vec k,0)&=&1 \hspace{1cm} \dot U_h(\vec k,0)
=-i\omega_{h0}(\vec k)\nonumber\\
U^*_h(\vec k,0)&=&1 \hspace{1cm}
\dot U^*_h(\vec k,0)=i\omega_{h0}(\vec k) \pkt
\end{eqnarray}
For $t = t'$ we get
\begin{equation}
G_{hh}^{++}(\vec k,t,t')= \frac{i}{2\omega_{h0}(\vec k)}
|U_h(\vec k,t)|^2 \pkt
\end{equation}
For the transversal components of the gauge fields
the mode functions satisfy
\begin{equation}
\left[\frac{\rm d^2}{\rm dt^2}+
\omega_\bot^2(\vec k,t)\right]U_\bot(\vec k,t)=0 
\end{equation}
where 
\be
\omega_\bot^2(\vec k,t)=\vec k^2+m_W^2+\frac{g^2}{4}
(H_0^2(t)-v^2)\pkt
\ee
The Green function is then given by
\begin{equation}
{G}^{++}_{\bot\bot}(\vec k,t,t)= \frac{i}{2\omega_{\bot0}(\vec k)}
|U_\bot(\vec k,t)|^2 
\end{equation}
where again $\omega_{\bot 0}$ is defined as
\be
\omega_{\bot0}(\vec k)=\omega_\bot(\vec k,0)\pkt
\ee

The Green functions for the time component $a^0$ of the gauge
field and of the isovector part of the Higgs field $\varphi$
satisfy the coupled system of differential equations
\begin{equation}
\left[(\frac{\rm d^2}{\rm dt^2}+\vec k^2+m_W^2)g_{mj}+
W_{mj}(t)\right]G_{jn}^{++}(\vec k,t,t')=\delta(t-t')\delta_{mn} \pkt
\end{equation}
Here we have introduced the metric ${\bf g}={\rm diag}\{-1,1\}$
taking into account the minus sign of the kinetic term and
of the propagator of $a_0$ in the Feynman gauge and
where the matrix ${\bf W}(t)$ is defined as
\begin{eqnarray}
\label{pot}
{\bf W}(t)
=\left(\begin{array}{cc}-
\frac{g^2}{4}(H_0^2(t)-v^2)&g\partial_0 H_0(t)\\
g\partial_0 H_0(t)&(\lambda+\frac{g^2}{4})(H_0^2(t)-v^2)
\end{array}\right)\pkt
\end{eqnarray}
The latin subscripts  $j,m,n$ take the values $a$ and $\varphi$.

For the Green functions of this  system we make the
Ansatz \cite{Baa,BaaSue}
\bea
G_{jn}^{++}(\vec k,t,t')&=&U_j^\alpha(\vec k,t)c_{\alpha\beta}(\vec k)
U_n^{\beta*}(\vec k,t,t')
\Theta(t-t')\nonumber\\&&+
U_j^{\alpha*}(\vec k,t)c_{\beta\alpha}(\vec k)
U_n^\beta(\vec k,t')\Theta(t'-t) \pkt
\eea
Here the mode equations
\begin{equation} \label{modeq}
\left[(\frac{\rm d^2}{\rm dt^2}+\vec k^2+m_W^2)g_{mj}+
W_{mj}(t)\right]U_{j}^{\alpha }(\vec k,t)=0
\end{equation}
have a fundamental system of two independent solutions
labelled by the superscript $\alpha$ which takes again
the values $a$ and $\varphi$. We choose as a
basis the two independent solutions characterized by the
initial conditions
\bea
\label{incon}
U^\alpha_j(\vec k,0)&=&\delta^\alpha_j \nonumber\\
\dot{U}^\alpha_j(\vec k,0)&=&-i\delta^\alpha_j \omega_{j0}(\vec k)
\end{eqnarray}
where
\be
\omega_{j0}(\vec k)=\left[\vec k^2+m_{j0}^2\right]^\frac{1}{2}
\ee
with
\bea \label{ma0}
m_{a0}^2 &=& m_{W0}^2 =m_W^2 + \frac{g^2}{4}(H_0^2(0)-v^2) \\
\label{mphi0}
m_{\varphi 0}^2&=& m_W^2 +(\lambda+\frac{g^2}{4})(H_0^2(0)-v^2)
\pkt
\eea
By a simple extension of the derivation given in
\cite{BaaSue} one can show that the 
coefficients $\cab(\vec k)$ are then related to the Wronskians
\bea
W(U^{\alpha},U^{\beta*},\vec k)&=&g_{mn}
\left(U^{\alpha}_m (\vec k,t)
\dot{U}^{\beta*}_n (\vec k,t)\right .\nonumber \\
&&\left .-\dot{U}_m^{\alpha}(\vec k,t) U_n^{\beta*}(\vec k,t)\right)
\eea
via
\be
\cab(\vec k) W(U^{\alpha},U^{\gamma*},\vec k)
 = - \delta^{\gamma}_{\beta} \pkt
\ee
The Wronskians can be computed from the initial conditions for
the $U^{\alpha }$; we obtain for the coefficients $\cab(\vec k)$
\be
\nonumber c_{\alpha\beta}(\vec k) = g_{\alpha\beta}/
(2\omega_{\alpha0}(\vec k))\pkt
\ee
We use the notation $U_j^\alpha$ for the single channel fields
$h$ and $\bot$ as well, implying that $U_h^h(\vec k,t)\equiv U_h(
\vec k,t) $
and $U_\bot^\bot(\vec k,t)\equiv
U_\bot(\vec k,t) $ and extend the metric 
to these components  via ${\bf g}={\rm diag }
\{1,1,-1,1\}$ corresponding to the ordering $h,\bot,a,\varphi$.
Collecting the various expressions for the Green functions
we define the (yet unrenormalized) fluctuation integral
\be
\label{fluctu}
{\cal F}(t) =
\sum_{jl\alpha}Q_{jl}(t)
\intk g_{\alpha\beta}
\frac{U_l^\alpha(\vec k,t)U_j^{\beta *}(\vec k,t)}{2
\omega_{\alpha 0}(\vec k)}     \pkt
\ee
 The equation of motion for $H_0(t)$ is then
\be
\label{motion}
\ddot{H}_0(t)+\lambda(H_0^2(t)-v^2)H_0(t) + {\cal F}(t) = 0 \pkt
\ee
In order to obtain the energy we first obtain the Hamiltonian from the
Lagrangean (\ref{freel}), insert the field expansion in terms of the
mode functions and the corresponding annihilation and creation
operators and take the expectation value in the initial state.
For the $a,\varphi$ subsystem we expand the fields
as 
\be
\left\{ \begin{array}{c} a(t,\vec x)\\ \varphi(t,\vec x)
\end{array}\right\}
=\intk \sum_\alpha\frac{1}{2\omega_{\alpha 0}(\vec k)}
\left[ c_\alpha(\vec k)\left\{\begin{array}{c} U_a^\alpha(t)\\ 
U_\varphi^\alpha(t)
\end{array}\right\}e^{i\vec k \vec x}+
c^\dagger_\alpha(\vec k)\left\{\begin{array}{c} U_a^{\alpha*}(t) 
\\ U_{\alpha*}^\varphi(t)
\end{array}\right\} e^{-i\vec k \vec x}\right]\pkt
\ee
With our initial conditions (\ref{incon})
for the mode functions $c_a,c_a^\dagger$ 
are at $t=0$ the annihilation and creation operators of
the field $a^0$ and $c_\varphi,c_\varphi^\dagger$ those of the
field $\varphi$. So averaging in the initial vacuum  state
of the system  we have in the Feynman gauge
 $<c_\alpha(\vec k) c_\beta^\dagger(\vec k')>_0 =
g_{\alpha\beta}(2\pi)^3 2
\omega_{\alpha0}(\vec k)\delta^3(\vec k -\vec k')
$. Therefore, the metric enters twice: once in order
to take into account the sign of the
kinetic terms of the field components (latin subscripts) 
in the Hamiltonian
 and a second time
due to the averaging of the field operators 
(greek indices) in the initial state. 
The unrenormalized
total energy of the system consisting of the inflaton field
and the fluctuations is therefore given by
\begin{eqnarray}
\label{energie}
{\cal E}&=&\frac{1}{2}\dot{H}_0^2(t)-
\frac{1}{4} m^2 H_0^2(t)+\pot{4}H_0^4(t)
\\\nonumber
&&+\intk\sum_{j,\alpha}g_{jj}g_{\alpha\alpha}
\frac{d_j}{2\omega_{\alpha 0}(\vec k)}
\left(\frac{1}{2}|\dot{U}_j^\alpha(\vec k,t)|^2
+\frac{\omega_j^2(\vec k,t)}{2}|U_j^\alpha(\vec k,t)|^2\right)
\end{eqnarray}
where the summation is over all fields $h,\bot,a$ and
$\varphi$.
 $d_j$ is the isospin degeneracy, i.e.  
$d_j=3$ for $j=\bot,a,\varphi$ and $d_j=1$ for $j=h$ \footnote{The
vertex operators are defined such to include such degeneracy factors.
The vertex operators $Q_{jl}$ and the potentials $V_{jl}$ are 
obviously related by
a functional derivative via 
$d_j\delta[\int dt'V_{jl} f(t')] /\delta H_0(t) = Q_{jl} f(t)$.}. 
The frequency $\omega_\alpha(t)$ is defined by
\be
\omega_\alpha^2(t)=\vec k^2 + m_\alpha^2(t) \pkt
\ee
Using the equations of motion for $H_0$ and the mode functions it can
be checked easily that the energy is conserved.
In both the fluctuation integral and the energy we have already taken
into account the cancellation
between the transversal gauge modes and the
Faddeev-Popov ghosts so that the latter ones do not appear any more.

\section{Perturbative expansion} \label{pertex}
In order to prepare the renormalized version of the
equations given in the previous section we introduce a suitable
expansion of the
mode functions $U_j^\alpha(t)$.
For a single channel this expansion
has been presented in \cite{BHP}.
We extend here the discussion to coupled channel systems.
Adding the term $(W_{mj}(t)-W_{mj}(0))U_j^\alpha(t)$ 
on both sides of the mode equations (\ref{modeq}) they take
the form
\begin{equation}
\label{udgl}
g_{jn}\left[ \frac{\rm d^2}{\rm dt^2}+
\omega^2_{n0}(\vec k)\right]U_n^\alpha(\vec k,t)=-V_{jl}(t)
U_l^\alpha(\vec k,t)
\end{equation}
with
\bea
{\bf V(t)}&=&{\bf W}(t)-{\bf W}(0) \\
&=&
\left(\begin{array}{cc}-\frac{g^2}{4}(H_0^2(t)-H_0^2(0))
&g\partial_0 H_0(t)\\
g\partial_0 H_0(t)&(\lambda+\frac{g^2}{4})(H_0^2(t)-H_0^2(0))
\end{array}\right) \nonumber
\eea
for the coupled system and
\bea
V_{hh}(t)&=&3\lambda(H_0^2(t)-H_0^2(0))\\
V_{\bot\bot}(t)&=&\frac{g^2}{4}(H_0^2(t)-H_0^2(0))
\eea
for the single component systems. We assume that ${\rm d}
H_0(t)/{\rm dt}$ to vanish at $t=0$.
Including the initial conditions (\ref{incon}) 
the mode functions satisfy the 
equivaltent integral equation
\begin{equation}
U_j^\alpha(\vec k,t)=\delta^\alpha_j e^{-i\omega_{j0} t}+
\int\limits^{\infty}_{0}\!{\rm d}t'
\Delta_{j,{\rm ret}}(\vec k,t-t')g_{jn}V_{nl}(t')U_l^\alpha(\vec k,t')
\end{equation}
with
\begin{equation}
\label{fvt}
\Delta_{j,{\rm ret}}(\vec k,t-t')=
-\frac{1}{\omega_{j0}}\Theta(t-t')
\sin(\omega_{j0}(t-t')) \; .
\end{equation}
We separate $U_j^\alpha(\vec k,t)$
into the trivial part corresponding to
the case $V_{jl}(t)\equiv 0$ and a function $f_j^\alpha(\vec k,t)$ 
which represents the
reaction to the potential by making the Ansatz
\begin{equation}
\label{ansatz}
U_j^\alpha(\vec k,t)=e^{-i\omega_{j0} t}
(\delta_j^\alpha+f_j^\alpha(\vec k,t)) \; .
\end{equation}
$f_j^\alpha(\vec k,t)$ satisfies then the integral equation
\begin{equation}\label{finteq}
f_j^\alpha(\vec k,t)=\int\limits^{t}_{0}\!{\rm d}t'
\Delta_{j,{\rm ret}}(\vec k,t-t')e^{i\omega_{j0} t}g_{jn}V_{nl}(t')
e^{-i \omega_{l0} t'}
(\delta^\alpha_l+f_l^\alpha(\vec k,t'))
\end{equation}
and an equivalent differential equation
\begin{equation}\label{fdiffeq}
(\dtt-2i\omega_{j0}(\vec k)\dt)f_j^\alpha(\vec k,t)=
-g_{jn}V_{nl}(t)
e^{i(\omega_{j0}-\omega_{l0})t}(\delta^\alpha_l
+f_l^\alpha(\vec k,t))
\end{equation}
with the initial conditions $f_j^\alpha(\vec k,0)=
\dot{f}_j^\alpha(\vec k,0)=0$.
\\ \\
We expand now $f_j^\alpha(\vec k,t)$
with respect to orders in $V_{jl}(t)$
by writing
\begin{eqnarray}
\label{entwicklung}
f_j^\alpha(\vec k,t)&=&
f_j^{(1)\alpha}(\vec k,t)+f_j^{(2)\alpha}(\vec k,t)+
f_j^{(3)\alpha}(\vec k,t) + .... \\
 &=& f_j^{(1)\alpha}(\vec k,t)+f_j^{\overline{(2)\alpha}}(\vec k,t)
\end{eqnarray}
where $f_j^{(n)\alpha}(\vec k,t)$ is of n'th order in $V_{jl}(t)$ and
$f_j^{\overline{(n)}\alpha}(\vec k,t)$
is the sum over all orders beginning with the n'th one. 
The $f_j^{(n)\alpha}(\vec k,t)$ are obtained by iterating the integral
equation (\ref{finteq}) or the differential equation
(\ref{fdiffeq}). The function $f_j^{\overline{(1)}\alpha}(\vec k,t)$ is
identical to the
function $f_j^\alpha(\vec k,t)$ itself which is obtained
by solving (\ref{fdiffeq}), the function
$f_j^{\overline{(2)}\alpha}(\vec k,t)$ can be obtained
as
\begin{equation}\label{f2inteq}
f_j^{\overline{(2)}\alpha}(\vec k,t)=
\int\limits^{t}_{0}\!{\rm d}t'\Delta_{j,{\rm ret}}(\vec k,t-t')
e^{i\omega_{j0} t} g_{jn}V_{nl}(t')e^{-i\omega_{l0} t'}
f_l^{\overline{(1)}\alpha}(\vec k,t')
\end{equation}
or by solving the inhomogeneous differential equation
\begin{equation}\label{f2diffeq}
(\dtt-2i\omega_{j0}(\vec k)\dt)
f_j^{\overline{(2)}\alpha}(\vec k,t)=-
g_{jn}V_{nl}(t)e^{i(\omega_{j0}-\omega_{l0})t}
f_j^{\overline{(1)}\alpha}(\vec k,t) \; .
\end{equation}
Note that in this way one avoids the computation 
of $f_j^{\overline{(2)}\alpha}(\vec k,t)$
via the small difference
$f_j^{\overline{(1)}\alpha}(\vec k,t)-f_j^{(1)\alpha}(\vec k,t)$. 
This feature is
especially important if deeper subtractions are required as in the
case of fermion fields. 

The order on the potentials $V_{jl}(t)$
will determine the behaviour of the functions
$f_j^{(n)\alpha}(\vec k,t)$ at large momentum.
We will give here the relevant leading terms for 
$f_j^{(1)\alpha}(\vec k,t)$ and
$f_j^{(2)\alpha}(\vec k,t)$. We have
\begin{equation}
f_j^{(1)\alpha}(\vec k,t)=\frac{i}{2 \omega_{j0}}
\int\limits^{t}_{0}\!{\rm d}t'
(e^{2 i \omega_{j0}(t-t')}-1)e^{i(\omega_{j0}-\omega_{\alpha 0}) t'}
g_{jn}V_{n\alpha}(t')  \; .
\end{equation}
Integrating by parts we obtain
\bea\label{f1exp}
f_j^{(1)\alpha}(\vec k,t)= 
&&-\frac{i}{2\omega_{j0}}\int\limits^{t}_{0}\!{\rm d}t'
g_{jn}V_{n\alpha}(t')e^{i(\omega_{j0}-\omega_{\alpha0})t'}
-\frac{1}{2\omega_{j0}(\omega_{j0}+\omega_{\alpha0})}
g_{jn}V_{n\alpha}(t)e^{i(\omega_{j0}
-\omega_{\alpha 0})t} \nonumber \\
&&+\frac{1}{2\omega_{j0}(\omega_{j0}+\omega_{\alpha0})}
\int\limits^{t}_{0}\!{\rm d}t'
e^{ 2i \omega_{j0} (t-t')}g_{jn}\dot{V}_{n\alpha}(t')
e^{i(\omega_{j0}-\omega_{\alpha0})t'} \pkt
\eea
For  $f_j^{(2)\alpha}(\vec k,t)$ we need to know only that the leading
behaviour is
\bea\label{f2exp}
f_j^{(2)\alpha}(\vec k,t)= 
&&-\sum_m\frac{1}{4\omega_{j0}\omega_{m0}}
\int\limits^{t}_{0}\!{\rm d}t'\int\limits^{t'}_{0}\!{\rm d}t''
g_{jl}V_{lm}(t')e^{i(\omega_{j0}-\omega_{m0})t'}
g_{mn}V_{n\alpha}(t'')e^{i(\omega_{n0}-\omega_{\alpha0})t''}
\nonumber \\&& + O((\omega_{ 0})^{-3}) \; .
\eea
The leading terms of 
$f_j^{\overline{(1)}\alpha}(\vec k,t)$ and 
$f_j^{\overline{(2)}\alpha}(\vec k,t)$
in this expansion in powers of $(\omega_{\alpha0})^{-1}$ are the same
as for $f_j^{(1)\alpha}(\vec k,t)$ and 
$f_j^{(2)\alpha}(\vec k,t)$ respectively.

\section{Renormalization} \label{renorm}
Using the expansion in orders of the potential $V_{jl}$
the fluctuation term (\ref{fluctu}) occuring in the 
equation of motion
(\ref{motion}) can be written as
\begin{equation}
\label{u}
{\cal F} (t) 
=\sum_{jl\alpha} Q_{jl}(t) g_{\alpha\alpha}
\intk \frac{(\delta^\alpha_l+f^\alpha_l(\vec k,t))
(\delta^\alpha_j+f^{\alpha*}_j(\vec k,t))}{2\omega_{\alpha0}}
e^{i(\omega_{j0}
-\omega_{l0})t}
\ee
To zeroth order in $V_{jl}$ the functions $f_j^\alpha$ vanish
and we have
\be
\label{tadpole}
{\cal F}^{(0)}(t)=\sum_l Q_{ll}(t)g_{ll}\intk\frac{1}{2\omega_{l0}}\pkt
\ee
These correspond to the tadpole graphs
depicted in Fig. 2; they are removed by 
including into the Lagrangean a mass counterterm 
for the Higgs field.

To first order in the potentials $V_{jl}$ we find
\bea
\label{fish}
&&{\cal F}^{(1)}(t) =
\sum_l Q_{ll}(t)g_{ll}\intk \frac{1}{2\omega^{0l}} 2 
\re f_l^{(1)l}(\vec k,t) \\
\nonumber
&&+Q_{a\varphi}(t) \intk \left\{
\frac{g_{\varphi\varphi}}{2\omega_{\varphi 0}} 2 
\re\left[ f_a^{(1)\varphi}(\vec k,t)e^{i(\omega_{\varphi 0}
-\omega_{a0}) t}\right ]-\frac{g_{aa}}{2\omega_{a0}} 2 
\re \left[f_\varphi^{(1)a}(\vec k,t)e^{i(\omega_{a0}-
\omega_{\varphi0})t}
\right]\right\}                    \pkt
\eea
The first part, the sum over the diagonal terms proportional
to $Q_{ll}(t)$,
corresponds to the graphs of Fig. 3a and their divergent
parts are removed by the coupling constant renormalization;
the second term corresponds to the graph of Fig. 3b and 
its divergent part is removed by a 
wave function renormalization counter term.

The sum of all contributions of order higher than $1$ 
in the potential $V_{jl}(t)$ is finite.
It is given by
\bea
&&{\cal F}^{\overline{(2)}}(t)
=\sum_{\alpha l} Q_{ll}(t)g_{ll}\intk\frac{g_{\alpha\alpha}}
{2\omega_{\alpha0}}
\left\{\delta_l^\alpha 2\re f^{\overline{(2)}l}_l(\vec k,t)
+f_l^{\overline{(1)}\alpha}(\vec k,t)
f_l^{\overline{(1)}\alpha*}(\vec k,t)\right\}\nonumber\\
&&-\frac{3}{2}g\dt\intk\left\{
\frac{g_{aa}}{2\omega_{a0}}2\re\left[
e^{i(\oma-\omp)t}(f_\varphi^{\overline{(2)}a}(\vec k,t)
+f_a^{\overline{(1)}a*}(\vec k,t)
f_\varphi^{\overline{(1)}a}(\vec k,t))\right]\right.
\nonumber\\
&&\hspace{2.7cm}\left.+\frac{g_{\varphi\varphi}}
{2\omp} 2\re\left[
e^{i(\omp-\oma)t}(f_a^{\overline{(2)}\varphi}(\vec k,t)
+f_a^{\overline{(1)}\varphi}(\vec k,t)
f_\varphi^{\overline{(1)}\varphi*}(\vec k,t))
\right]\right\}                        \pkt\nonumber\\
\eea

The divergent zeroth and first order terms have now to be 
defined by a regularization and the divergent parts to be absorbed
into appropriate counter terms. In \cite{BHP} we found
dimensional regularization to be suitable both for Lorentz
covariance and for technical elegance.
The zeroth order term can be handled as in \cite{BHP} via
\bea
{\cal F}^{(0)}_{\rm reg}(t)
&=&\sum_l g_{ll}Q_{ll}(t) \mu^\epsilon
\int \frac{{\rm d}^{3-\epsilon}k}{(2\pi)^{3-\epsilon}}
\frac{1}{2\omega_{l0}} \nonumber \\
&=&- \sum_l \frac{g_{ll}Q_{ll}(t)m_{l0}^2}{16\pi^2} 
\left\{\frac{2}{\epsilon}+ \ln \frac{4 \pi \mu^2}{m_{l0}^2}-
\gamma + 1 \right \} \pkt
\eea
The divergent part of this expression depends on the
`initial masses' $m_{l0}$, the discussion of the counterterm
is postponed, therefore, until we have discussed the first order
term as well. The first order part consists of  a sum
over diagonal terms proportional to $Q_{ll}$ and a 
nondiagonal part containing $Q_{a\varphi}$.

The diagonal part can be handled as in \cite{BHP}; using
(\ref{f1exp}) we obtain
\bea
\label{dimreg2}
&&\sum_l g_{ll}Q_{ll}(t)\intk \frac{1}{2\omega_{l0}}
 2 \re f_l^{(1)l}(\vec k,t) 
= \sum_l \frac{Q_{ll}(t) \mu^\epsilon }
{16\pi^2} V_{ll}(t)\left\{\frac
{2}{\epsilon}
+\ln{\frac{4\pi\mu^2}{m_{0l}^2}}-\gamma\right\} \nonumber\\
&&+ \sum_lQ_{ll}(t)
\intk\frac{1}{4 \omega^3_{l0}}\int_0^t {\rm d}t'
\cos(2\omega_{l0}(t-t'))\dot{V}_{ll}(t') \pkt 
\eea
Again using the expansion (\ref{f1exp}) 
we obtain for the nondiagonal part
\bea
&&Q_{a\varphi}(t) \intk \left\{
\frac{g_{\varphi\varphi}}{2\omega_{\varphi0}} 2 
\re\left[ f_a^{(1)\varphi}(\vec k,t)e^{i(\omega_{\varphi0}
-\omega_{a0}) t}\right ]+\frac{g_{aa}}{2\omega_{a0}} 2 
\re \left[f_\varphi^{(1)a}(\vec k,t)
e^{i(\omega_{a0}-\omega_{\varphi0})t}
\right]\right\} \nonumber
\\  
&&=-\frac{3}{2}g^2 
\ddot{H}_0(t)\intk\frac{g_{aa}g_{\varphi\varphi}} 
{\oma\omp(\oma+\omp)}\\
&&+\frac{3}{2}g^2 
\partial_t\intk\frac{g_{aa}g_{\varphi\varphi}}{\oma\omp(\oma+\omp)}
\int\limits_{0}^t\!{\rm d} t'\, 
\ddot{H}_0(t')\cos((\oma+\omp)(t-t')) \nonumber \pkt
\eea
The first integral on the right hand side is divergent; 
the connection to 
the wave function renormalization in
usual time-ordered perturbation theory
 becomes transparent if we rewrite it
as
\be
\intk\frac{1}{\oma\omp(\oma+\omp)}
= \int\frac{{\rm d}^4k}{(2\pi)^4}\frac{1}
{(k^2-m_{0a}^2+io)(k^2-m_{0\varphi}^2+io)} \pkt
\ee
Using dimensional regularization we find
\bea
&&\int\!\frac{{\rm d}^{3-\epsilon}k}{(2\pi)^{3-\epsilon}}\,
\frac{1}{\oma\omp(\oma+\omp)} \\ =
&&\frac{1}{16\pi^2}
\left\{\frac{2}{\epsilon}-\gamma +1
+\ln\frac{4\pi\mu^2}{m_{a0}^2}+\frac{m_{\varphi 0}^2}
{m_{a0}^2-m_{\varphi 0}^2}\ln \frac{m_{\varphi 0}^2}{m_{a0}^2}
\right \}\nonumber \pkt
\eea

As in \cite{BHP} there are cancellations between the zeroth order
and first order divergencies such that the counter terms can be
chosen independent of the initial conditions, i. e. they can
be written in terms of the 
masses $ m_l$ rather than the masses `at time zero', $m_{l0}$.
We renormailze at $q^2=0$ and chose the counter terms in such a
way that the corrections to the tree effective potential 
vanish at its minimum $|\Phi|=H_0=v=m_h^2/2\lambda$. 
In particular we find the wave function renormalization counter term
\begin{equation}
\delta Z=
\frac{3g^2}{32\pi^2}\left\{\frac{2}{\epsilon}
+\ln{\frac{4\pi\mu^2}{m_W^2}}-\gamma+1
\right\}\pkt
\ee
The mass renormalization counter terms takes the form
\bea
\delta m_h^2
&=&\frac{3}{16\pi^2}\lambda m_h^2 \left\{\frac{2}{\epsilon}
+\ln{\frac{4\pi\mu^2}{m_{h}^2}}-\gamma+1\right\}\nonumber\\
&&+\frac{3}{16\pi^2}\left(\lambda+\frac{g^2}{4}\right)m_h^2 
\left\{\frac{2}{\epsilon}
+\ln{\frac{4\pi\mu^2}{m_h^2}}-\gamma+1\right\}\pkt
\end{eqnarray}
Finally the coupling is renormalized via
\bea
\delta\lambda
&=&\frac{9}{16\pi^2}\lambda^2\left\{\frac{2}{\epsilon}
+\ln{\frac{4\pi\mu^2}{m_h^2}}-\gamma\right\}\nonumber\\
&&+\frac{3}{128\pi^2}g^4\left\{\frac{2}{\epsilon}
+\ln{\frac{4\pi\mu^2}{m_a^2}}-\gamma\right\}\nonumber\\
&&+\frac{3}{16\pi^2}\left(\lambda+\frac{g^2}{4}\right)^2
\left\{\frac{2}{\epsilon}
+\ln{\frac{4\pi\mu^2}{m_h^2}}-\gamma\right\}\pkt
\end{eqnarray}
This choice of the counter terms corresponds to a renormalization
of the Green functions at $q^2=0$ as usual in the renormalization
of the effective potential.
If these counter terms are introduced into the
 equation of motion, all divergencies of ${\cal F}$ are cancelled, but
 we are still left with the respective finite parts
\begin{eqnarray}
\Delta Z&=&-\frac{3g^2}{32\pi^2}\left\{\frac{m_{\varphi 0}^2}
{m_{a 0}^2-m^2_{\varphi 0}}\ln{\frac{m_{\varphi 0}^2}{m_{a 0}^2}}
+\ln{\frac{m_a^2}{m_{a 0}^2}}\right\}\\
\Delta m_h^2&=&-\frac{3}{16\pi^2}
\lambda m_h^2\ln{\frac{m_h^2}{m_{h 0}^2}}
+\frac{9}{8\pi^2}\lambda^2 H^2_0(0)\nonumber\\
&&+\frac{3}{16\pi^2}g^2m^2_{a 0} H_0^2(0)\nonumber\\
&&-\frac{3}{16\pi^2}(\lambda+\frac{g^2}{4})m_h^2\ln{\frac{m_h^2}
{m_{\varphi0}^2}}+\frac{3}{8\pi^2}(\lambda+\frac{g^2}{4})^2 H^2_0(0)\\
\Delta \lambda&=&
-\frac{9}{16\pi^2}\lambda^2 \ln{\frac{m_h^2}{m_{h 0}^2}}
-\frac{3}{128\pi^2}g^4\ln{\frac{m_a^2}{m_{a
0}^2}}\\
&&-\frac{3}{16\pi^2}(\lambda+\frac{g^2}{4})^2\ln{\frac{m_h^2}
{m_{\varphi 0}^2}}\nonumber\pkt
\end{eqnarray}
With these definitions the equation of motion reads
\begin{equation}
(1+\Delta Z_{H_0})\ddot{H}_0(t)-\frac{1}{2}(m^2_h+\Delta m^2_h)H_0(t)
+(\lambda+\Delta\lambda)H^3_0(t)+{\cal F}_{\rm fin}(t)=0,
\end{equation}     
where the finite part of the fluctuation integral is given by
\bea
{\cal F}_{\rm fin}(t) &=&\intk
\left\{\sum_l Q_{ll}(t)
\frac{g_{ll}}{4\omega^3_{l0}}\int\limits^{t}_{0}\!{\rm
d}t'\cos(2\omega_{l0}(t-t'))\dot{V}_{ll}(t')\nonumber\right.\\
&&+\sum_{l\alpha}Q_{ll}(t)\frac{g_{\alpha\alpha}}
{2 \omega_{\alpha0}}\left\{2
\delta_l^\alpha{\rm Re}f_l^{\overline {(2)}l}(\vec k,t)
+f_l^{\overline {(1)}\alpha}(\vec k,t)
f_l^{\overline {(1)}\alpha*}(\vec k,t)\right\}\\
&&+\frac{3}{2}g^2\partial_t\frac{1}{\oma\omp(\oma+\omp)}
\int\limits_{0}^t\!{\rm d} t'\, 
\ddot{H}_0(t')\cos((\oma+\omp)(t-t'))\nonumber\\
&&-\frac{3}{2}g\partial_t\left[\frac{g_{aa}}{\oma}
\re \left(e^{i(\oma-\omp)t}(
f_{\varphi}^{\overline{(2)}a}
(\vec k,t)+f_{\varphi}^{\overline{(1)}a}(\vec k,t)
f_{a}^{\overline{(1)}a*}(\vec k,t))\right)\right.\nonumber\\
&&\left.\left.\hspace{1.3cm}+\frac{g_{\varphi\varphi}}{\omp}\re
\left(e^{i(\omp-\oma)t}(
f_{a}^{\overline{(2)}\varphi}(\vec k,t)
+f_{a}^{\overline{(1)}\varphi}(\vec k,t)
f_{\varphi}^{\overline{(1)}\varphi*}(\vec k,t))\right)\right ]
\right\}\pkt \nonumber
\end{eqnarray}
The infinite and finite counter terms introduced so far 
have to be included into the expression for the energy as well.
In addition a new counter term is required to compensate the
quartically divergent zero point energy. It has again an infinite
part determined by the renormalization condition at $q^2=0$ and
$|\Phi|=v$
\be
\delta \Lambda = \frac{m_h^4}{64 \pi^2}\left\{
\frac{2}{\epsilon} +\ln \frac{4\pi \mu^2}{m_h^2}-\gamma+
\frac{3}{2} \right\}
\ee
for the divergent part and
\be
\Delta\Lambda=\frac{m_h^4}{256\pi^2}
\left\{\ln\frac{m_h^2}{m_{h0}^2} +3\ln\frac{m_h^2}{m_{\varphi 0}^2}
\right\} 
\ee
for the finite part. In addition the quadratically and logarithmically
terms contain some finite parts which depend on the initial
value of $H_0(0)$; they are constant and do not appear in the
equation of motion, therefore. They lead to another finite correction
to the energy
\be
\Delta\Lambda'=\frac{1}{128 \pi^2}\left\{
\left(9\lambda^2+\frac{3g^4}{8}+3\left(\lambda
+\frac{g^2}{4}\right)^2\right) H_0^4(0)
+\left(3\lambda+3\left(\lambda+
\frac{g^2}{4}\right)\right)H_0^2(0)\right\} \pkt
\ee
Altogether we obtain for the renormalized energy
\begin{eqnarray}
{\cal E}&=&\frac{1}{2}(1+\Delta Z_{H_0})\dot{H_0}^2(t)
-\frac{1}{4}(m^2_h+\Delta m^2_h) H_0^2(t)\nonumber\\
&&+\frac{\lambda+\Delta\lambda}{4}H_0^4(t)\nonumber
+\Delta\Lambda +\Delta \Lambda'\\
&&+
\intk \sum_{j\alpha}
 \frac{d_j g_{jj}g_{\alpha\alpha}}{2\omega_{\alpha0}}
\left\{\omega^2_{j0}(2\re
\delta_j^\alpha f_{j}^{\overline{(2)}\alpha}(\vec k,t)+
|f_{j}^{\overline{(1)}{\alpha}}(\vec k,t)|^2)\right.\\
&&\hspace{2.4cm}+\frac{1}{2}|\dot{f}_j^
{\overline{(1)}\alpha}(\vec k,t)|^2
-\omega_{j0}\re(i\delta_j^\alpha
\dot{f}_j^{\overline{(2)}\alpha*}(\vec k,t)
+if_j^{\overline{(1)}\alpha}(\vec k,t)
\dot{f}_j^{\overline{(1)}\alpha*}(\vec k,t))\nonumber\\
&&\hspace{2.4cm}\left.+\frac{V_{jj}(t)}{2}
(\delta_l^\alpha+2\delta_j^\alpha\re 
f_j^{\overline{(1)}\alpha}(\vec k,t)+
|f_j^{\overline{(1)}\alpha}(\vec k,t)|^2)+
\frac{V_{j\alpha}^2(t)}{4\omega_{j0}(\omega_{j0}+\omega_{\alpha0})}
\right\}\nonumber
\end{eqnarray}
We denote the sum of the first five terms as the "inflaton energy"
and the last term as "fluctuation energy". Of course the inflaton
energy includes besides the tree level energy the finite terms
left over after renormalization. In terms of potentials we 
can distinguish between:
\\ \\
the tree level potential (\ref{tree}), i. e.
\be
V_{\rm tree}(H_0) = \frac{1}{2}\mu^2 H_0^2+\frac{\lambda}{4}H_0^4,
\ee
the effective potential appearing in the inflaton energy
\be 
\tilde{V}_{\rm eff} =-\frac{1}{4}(m^2_h+\Delta m^2_h) H_0^2(t)
+\frac{\lambda+\Delta\lambda}{4}H_0^4(t)
+\Delta\Lambda +\Delta \Lambda'  
\ee
and
the renormalized one-loop potential
\bea
V_{\rm 1-l}(H_0)&=
&\frac{m_h(H_0)^4}{64\pi^2}\ln\frac{m_h^2}{m_h^2(H_0)}
+ \frac{3 m_a(H_0)^4}{32\pi^2}\ln\frac{m_W^2}{m_a^2(H_0)}
\nonumber \\
&&+\frac{3m_\varphi(H_0)^4}{64\pi^2}\ln\frac{m_h^2}{m_\varphi^2(H_0)}
+\frac{3H_0^4}{128\pi^2}\left(9\lambda^2+\frac{3}{8}g^4 
+3\left(\lambda+\frac{g^2}{4}\right)^2\right) \nonumber\\
&&-\frac{3}{128\pi^2}
m_h^2 H_0^2 \left(2\lambda+\frac{g^2}{4}\right) \pkt
\eea 
Here $m_h^2(H_0)=m_h^2+3 \lambda(H_0^2-v^2)$ while
$m_a(H_0)$ and $ m_\varphi(H_0)$ are given by
Eqns. (\ref{ma0},\ref{mphi0}) whith $H_0(0)$ replaced by $H_0$.

\section{Results and Conclusions}
\label{conclus}

The computation scheme presented in this paper can be 
implemented numerically along the lines described
in our previous work \cite{BHP} for the case of a
self-coupled scalar field. We have to deal here
with a coupled-channnel system which, however, can
be treated using perfectly analogous methods; there
is of course a noticeable increase in  CPU time.
If the Higgs mass is chosen as the mass scale  
the $SU(2)$ Higgs model has two parameters, $g$ and
$\lambda$, the initial value of the Higgs field is an
additional variable. We will not attempt here to 
exhaust this parameter space by presenting results
for a systematic choice of samples. Rather we restrict
ourselves here to two examples; a systematic study
should include the development of new concepts like the
`true effective potential' \cite{CoHaKlMo} or at least the
static one loop effective potential (see below). Especially
for gauge theories with their numerous degrees of freedom
one-loop corrections tend to be large and the use of
the tree potential and the bare parameters may  lead
to wrong interpretations of the results.

We have chosen two parameters sets. The first one
has a large initial value for the Higgs field leading,
in spite of the mexican-hat potential, to symmetric 
oscillations; the paramters are $g=0.1$, $\lambda=1$ and
$H_0(0)=10$. The results are presented in Figs. 4 - 6. The
classical amplitude decreases slowly with time, the energy is 
transferred almost entirely to the fluctuations.    

The second example is chosen in such a way that the 
inflaton field oscillates around its vacuum expectation value.
Here the parameters are: $g=1$, $\lambda=1$ and 
$H_0(0)=0.65$. We observe again a strong decrease of the classical
amplitude and a corresponding transfer of energy. 

For both parameter sets we present also the 
inflaton energy as a
function of $H_0$. For time-independent $H_0$ this energy
would be given by the sum of tree  and 
the renormalized one-loop effective
potential $V_{\rm tree}+V_{\rm 1-l}$. 
The actual motion is governed, however, by the potential 
$\tilde{V}_{\rm eff}$.
It is surprising that this potential 
which is determined entirely by the initial conditions 
governs the motion even after the quantum
fluctuations are fully developed.

In conclusion we have extended here our method for the numerical
study of nonequilibrium processes to the $SU(2)$ Higgs model
and thereby to a system which is typical for the inflationary
scenario. It was again possible to separate the numerical
computation of finite part of the fluctuation integrals
 and the analytic evaluation of their
renormalization parts. We have not studied here initial conditions
where the inflaton field starts out at the metastable 
maximum $H_0 =0$ or anywhere in the region where the 
effective potential has negative curvature or is even complex.
As we have mentioned above such a study would have to be
accompanied by a study of effective potentials.  
Such investigations could help to develop our understanding
of  negative curvature or
complex effective potentials, as analyzed theoretically
by Weinberg and Wu \cite{WeWu}.

\newpage
\section*{Figure captions}
\noindent
{\bf Fig. 1 } Diagrammatic representation of the 
one-loop equation (\ref{fdiffeq}).
\\ \\
{\bf Fig. 2a } Renormalization parts: Tadpole diagram,
Eq. (\ref{tadpole}).
\\ \\
{\bf Fig. 3a } Renormalization parts: Fish diagram ,
Eq. (\ref{fish}).
\\ \\
{\bf Fig. 3b } Renormalization parts: Fish diagram ,
Eq. (\ref{fish}).
\\ \\
{\bf Fig. 4} The inflaton amplitude $H_0(t)$ for 
$\lambda=1$, $g=0.1$ and $H_0(0) = 10$.
\\ \\
{\bf Fig. 5} The fluctuation integral $\Delta {\cal F}(t)$ 
for the same set of parameters.
\\ \\
{\bf Fig. 6} The inflaton and fluctuation energies as 
a function of time for the same parameters:
inflaton energy (short dashed line), the fluctuation energy
(long dashed line) and total energy (solid line) .
\\ \\
{\bf Fig. 7} The inflaton amplitude $H_0(t)$ for
$\lambda=1$, $g=1$ and $H_0(0) = 0.65$.
\\ \\
{\bf Fig. 8} The inflaton and fluctuation energies as 
a function of time for the same parameters:
inflaton energy (short dashed line), the fluctuation energy
(long dashed line) and total energy (solid line) .
\\ \\
{\bf Fig. 9} The inflaton energy as a function of $H_0(t)$
for the first parameter set; we also display
(a) the tree level potential $V_{\rm tree}$,
(b) the sum of tree level and one-loop effective potentials
and (c) the potential $\tilde{V}_{\rm eff}$ .
\\ \\
{\bf Fig. 10} The same as Fig. 9 for the second parameter set.

\newpage
\vspace*{10mm}
\parbox{10cm}
\mbox{\begin{center} \mbox{\epsfxsize=6cm\epsfbox{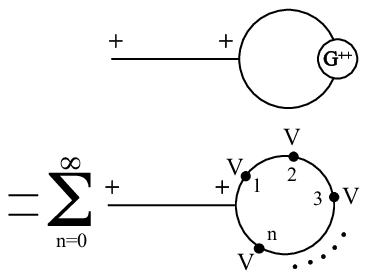}}
\end{center}}
\vspace{10mm}
\begin{center}
{\Large \bf Fig. 1}
\end{center}
\vspace*{10mm}
\parbox{10cm}
\mbox{\begin{center} \mbox{\epsfxsize=6cm\epsfbox{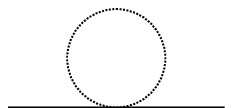}}
\end{center}}
\vspace{10mm}
\begin{center}
{\Large \bf Fig. 2}
\end{center}
\vspace*{10mm}
\parbox{10cm}
\mbox{\begin{center} \mbox{\epsfxsize=6cm\epsfbox{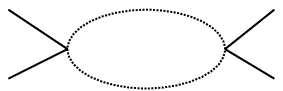}}
\end{center}}
\vspace{10mm}
\begin{center}
{\Large \bf Fig. 3a}
\end{center}
\vspace*{10mm}
\parbox{10cm}
\mbox{\begin{center} \mbox{\epsfxsize=6cm\epsfbox{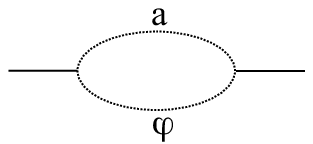}}
\end{center}}
\vspace{10mm}
\begin{center}
{\Large \bf Fig. 3b}
\end{center}

\newpage
\parbox{15cm}
\mbox{\begin{center} \mbox{\epsfxsize=15cm\epsfbox{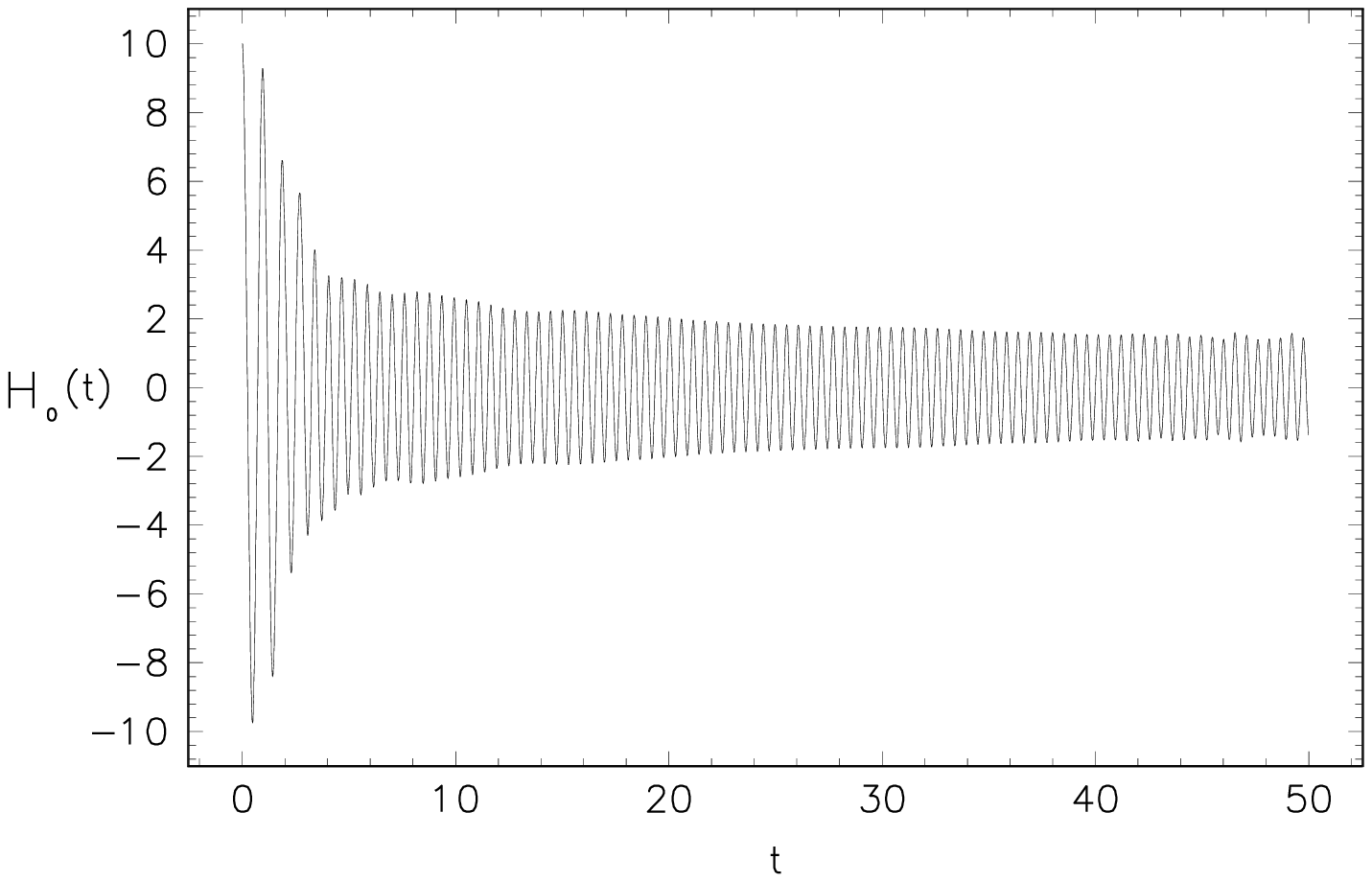}}
\end{center}}
\begin{center}
{\Large \bf Fig. 4}
\end{center}
\parbox{15cm}
\mbox{\begin{center} \mbox{\epsfxsize=15cm\epsfbox{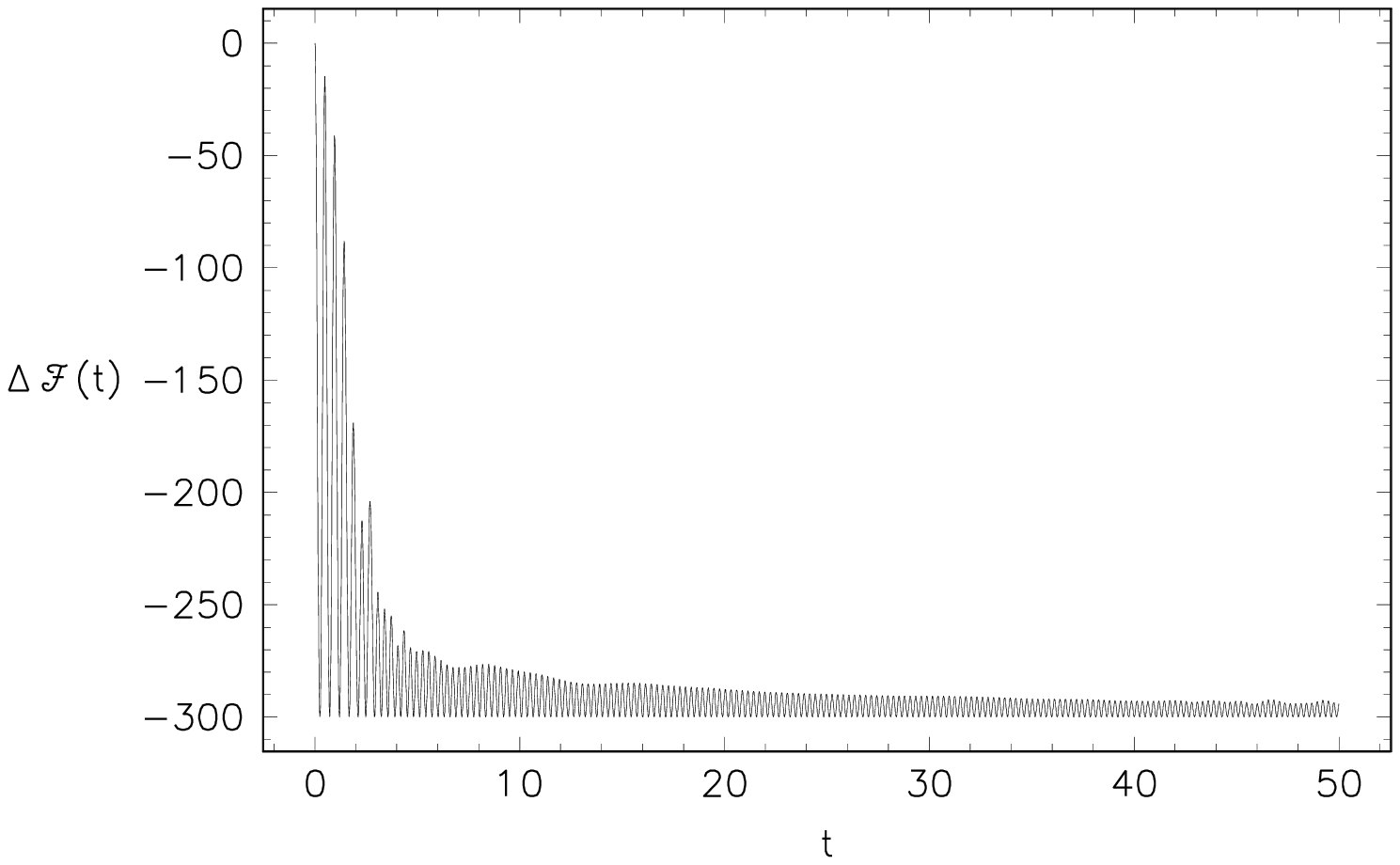}}
\end{center}}
\begin{center}
{\Large \bf Fig. 5}
\end{center}
\newpage
\parbox{15cm}
\mbox{\begin{center} \mbox{\epsfxsize=15cm\epsfbox{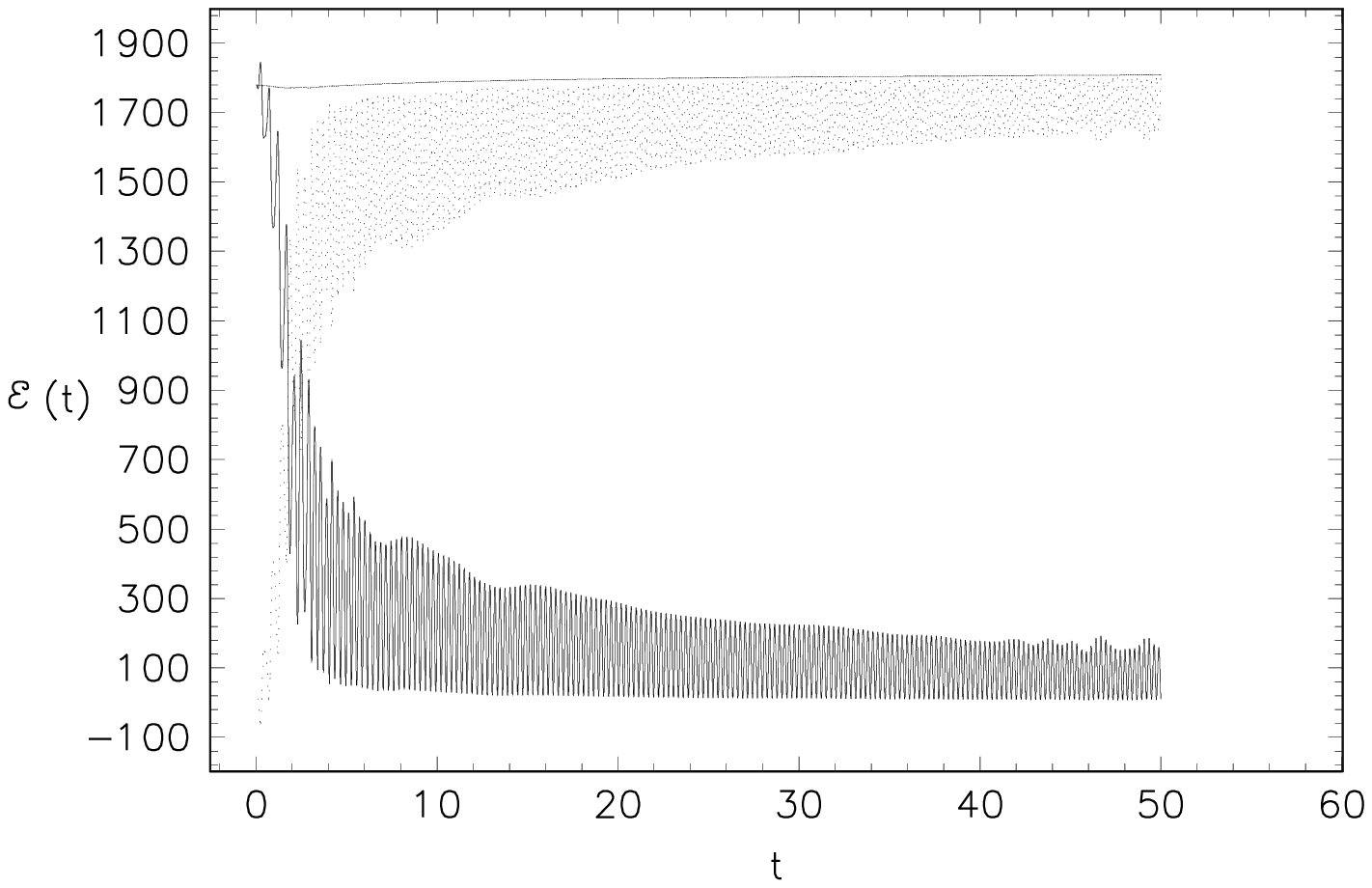}}
\end{center}}
\begin{center}
{\Large \bf Fig. 6}
\end{center}
\parbox{15cm}
\mbox{\begin{center} \mbox{\epsfxsize=15cm\epsfbox{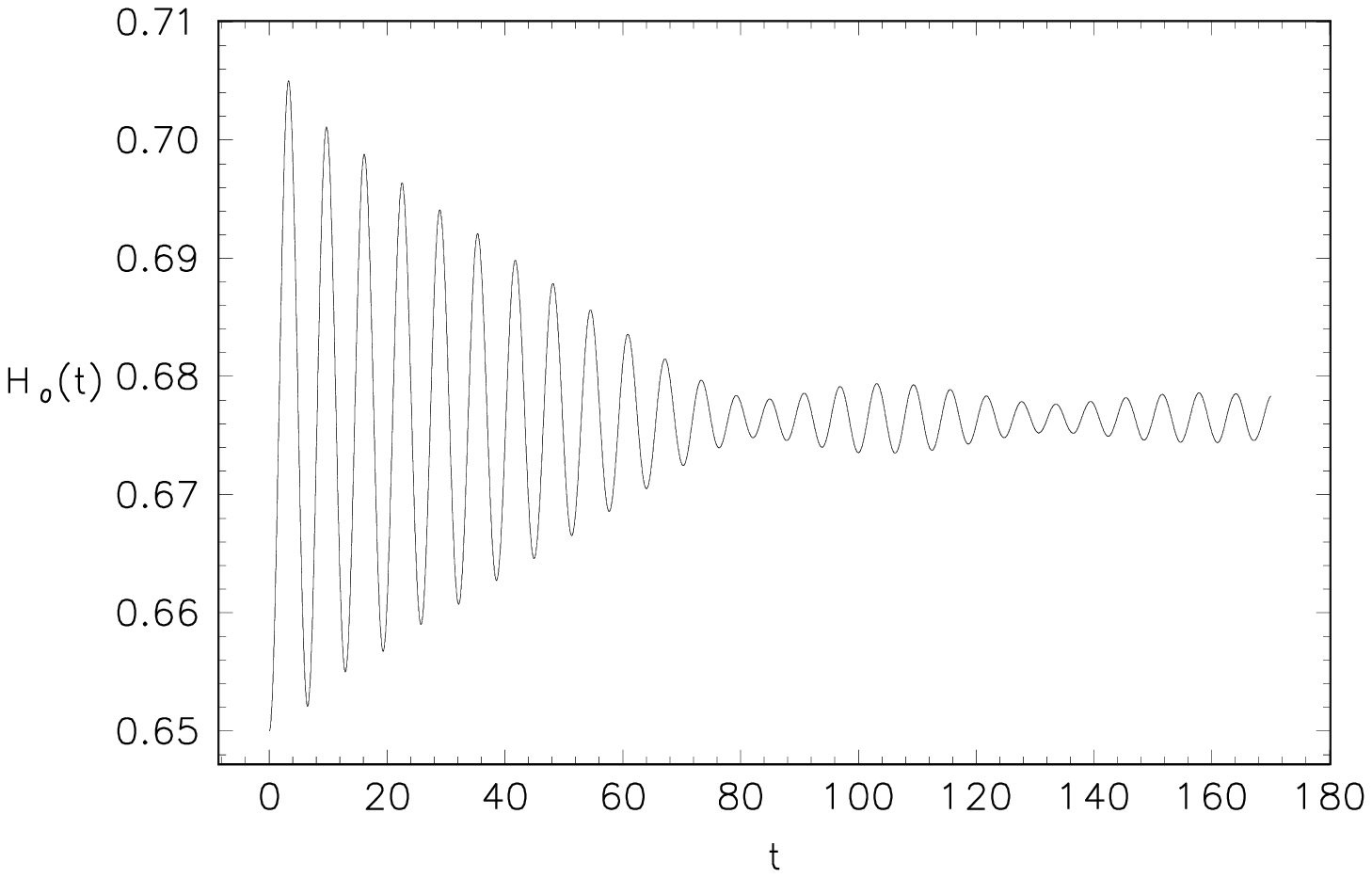}}
\end{center}}
\begin{center}
{\Large \bf Fig. 7}                                
\end{center} 
\parbox{15cm}
\mbox{\begin{center} \mbox{\epsfxsize=15cm\epsfbox{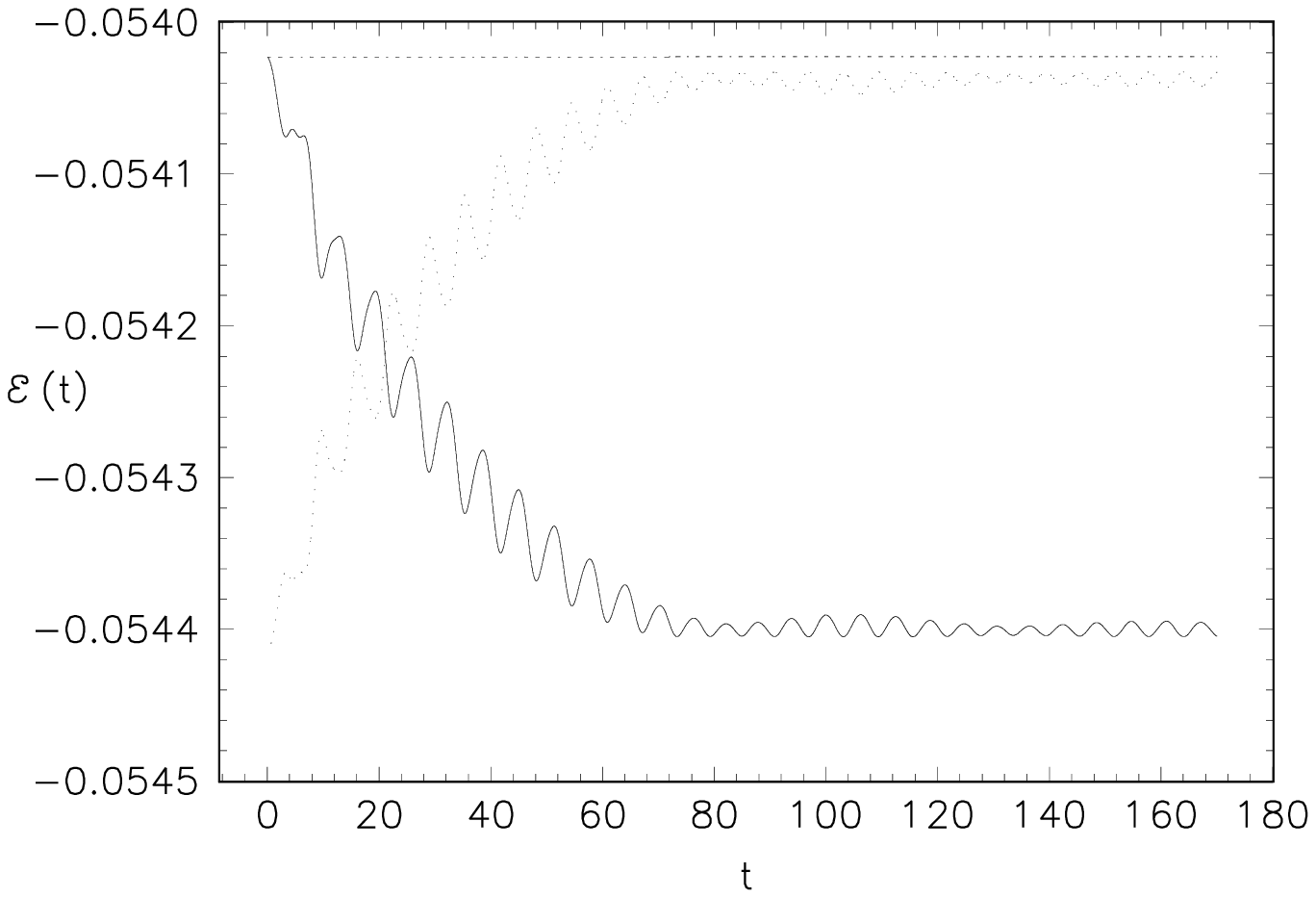}}
\end{center}}
\begin{center}
{\Large \bf Fig. 8}
\end{center}
\newpage
\parbox{15cm}
\mbox{\begin{center}\mbox{\epsfxsize=15cm\epsfbox{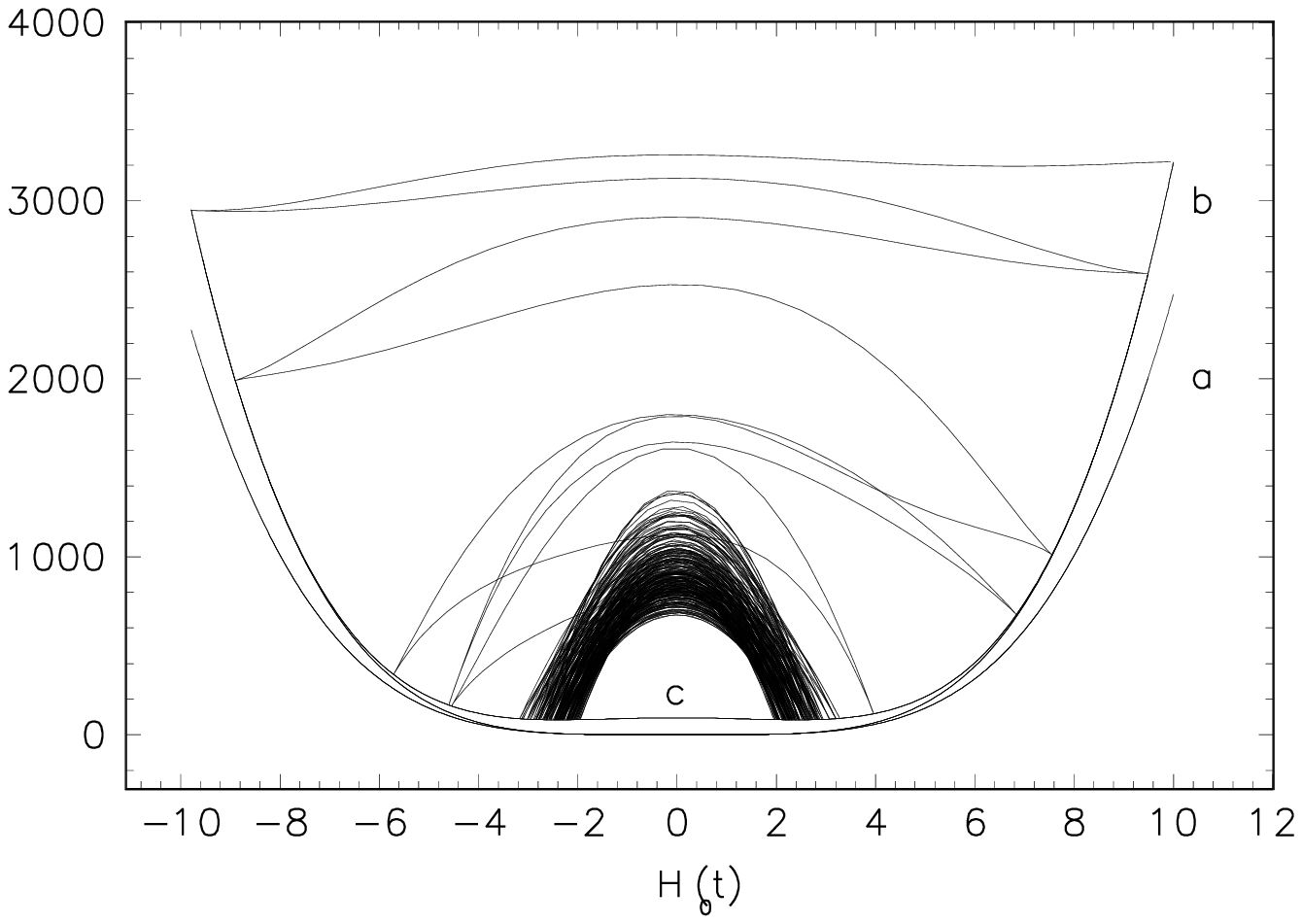}}
\end{center}}
\begin{center}
{\Large \bf Fig. 9}
\end{center}
\parbox{15cm}
\mbox{\begin{center}\mbox{\epsfxsize=15cm\epsfbox{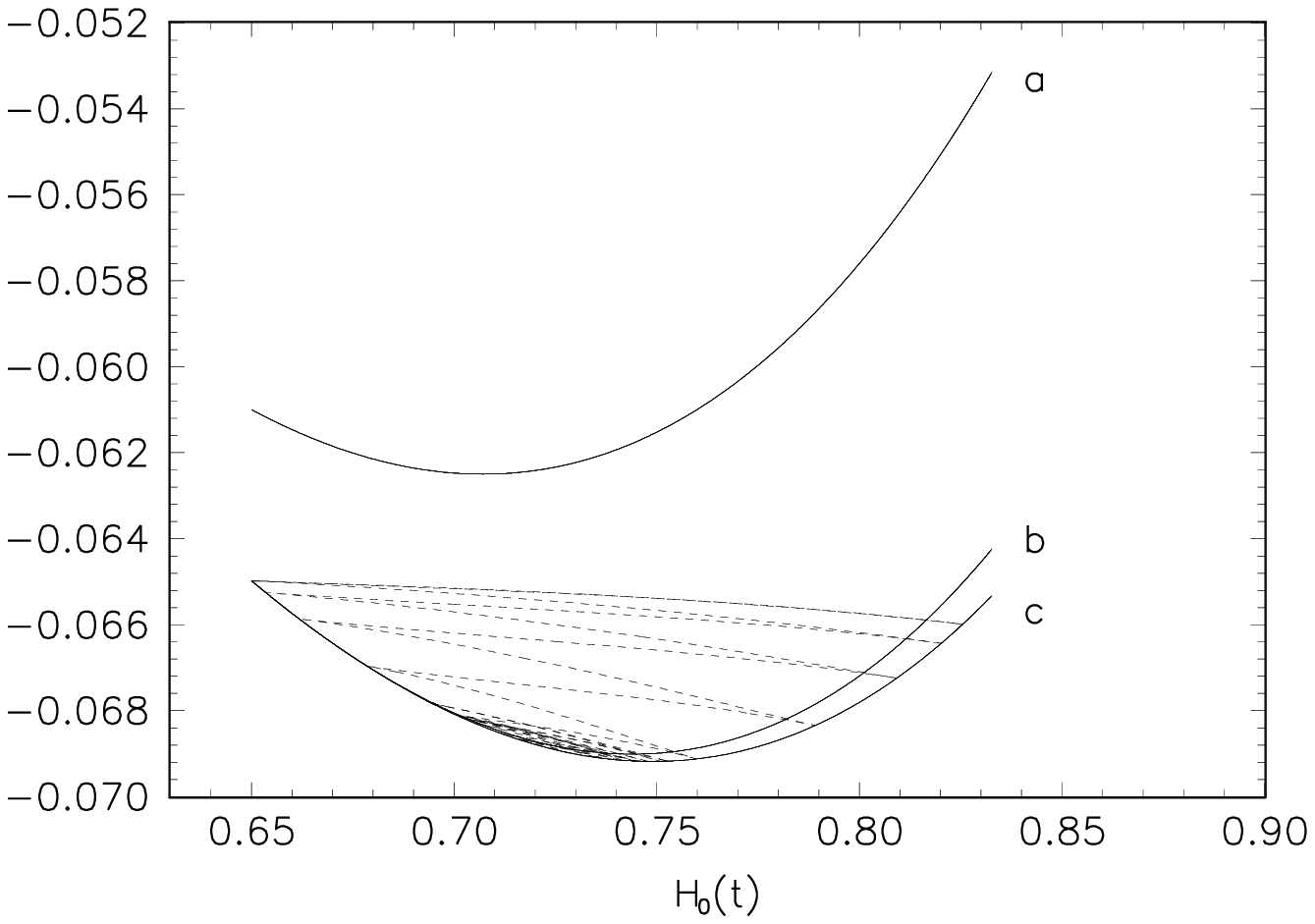}}
\end{center}}
\begin{center}
{\Large \bf Fig. 10}
\end{center}

\end{document}